\documentstyle[preprint,aps,eqsecnum]{revtex}

\begin{document}
\draft
\title{\bf
MAGNETIC MONOPOLES IN THE EINSTEIN-YANG-MILLS-HIGGS SYSTEM
}

\author{
{\bf Nguyen Ai Viet}$^*$ \hskip 3mm
{\bf and Kameshwar C.Wali}\\
{\normalsize  Physics Department, Syracuse University, Syracuse, New York
13244}}
\maketitle
\begin{abstract}
We study the Yang-Mills-Higgs system within the framework of general
relativity. In
the static situation, using Bogomol'nyi type analysis, we derive a
positive-definite energy functional which has a lower bound. Specializing to
the gauge group $SU(2)$ and the t'Hooft-Polyakov ansatz for the gauge and Higgs
fields, we seek static, spherically symmetric solutions to the coupled system
of equations in both the isotropic and standard coordinate systems. In both
cases, in the spontaneously broken symmetry situation, we find great
simplications reducing the solutions of the coupled system to the solution of a
single non-linear differential equation, different one in each case, but
well-known in other contexts of physics. We find abelian and non-abelian
monopole solutions with gravitational fields playing the role of Higgs fields
in providing attraction that balances the repulsion due to the gauge fields.
Numerical solutions indicate the possibility of blackhole horizons inside the
monopoles enclosing the singularity at the origin. Such non-abelian monopoles
are then the analogs of Reissner-Nordstr\"om blackholes with magnetic charge.
\end{abstract}


\section{Introduction}
Recently there has been considerable interest in the study of classical, static
and spherically symmetric solutions of coupled Einstein-Yang-Mills-Higgs [EYMH]
system [1]. The main purpose of such investigations has been to understand the
properties of non-abelian monopoles such as the t'Hooft-Polyakov monopoles in
the presence of gravity. The latter, which are finite energy configurations of
Yang-Mills and Higgs fields, exhibit extended structure and the Coulomb
interactions between two monopoles or a monopole and an anti-monopole are
altered by the complex interactions between the gauge fields and Higgs fields
[2]. The question that arises naturally is what happens when such monopoles are
thought to be in curved space-time described by a given background metric.

It is of some interest in this connection to note that, in the framework of
Newtonian mechanics, a system of $N$ mass points with like electrical
(or magnetic)
charges can remain in perfect static equilibrium if their masses are, in
suitable units, equal to their charges. Remarkably, a generalization of this
situation exists within the framework of general relativity described by the
Majumdar-Papapetrou solution [3], which was shown later to represent $N$
extreme Reissner-Nordstr\"om black holes [4]. One may, therefore, ask whether
there is a corresponding generalization of Yang-Mills-Higgs monopoles.
Nieuwenhuizen, Wilkinson and Perry [5] were the first ones to investigate
static
solutions in the case of EYMH system. They showed that one could construct a
positive-definite energy functional for the complete system, which at its
minimum, gave a regular solution for the t'Hooft-Polyakov monopole. More
recently, Bartnik and McKinnon found numerical solutions [6] of the
Einstein-Yang-Mills equations, which had particle-like properties and
demonstrated among other things that gravitational interaction could not be
dismissed as too weak to be consequential. Their solutions depended
non-trivially on the coupled Yang-Mills and Einstein fields. Bais and Russell,
and independently Cho and Freund found a solution [7] to the complete EYMH
system, which, however, has the abelian Dirac monopole character with a
singularity at the origin. One is interested in more general solutions with
non-abelian character and in the question; whether such solutions are stable
and
whether they have event horizons inside or outside the core defined by the
Yang-Mills-Higgs fields in the absence of gravity.

The present paper is an extension of the work by Balakrishna and Wali [8]. With
a specific coupling of the Higgs field to the Einstein term in the Lagrangian
describing the EYMH system, they showed that, in the static case, the energy
functional can be reduced, by a Bogomol'nyi-type [9] condition, to a form that
resembles the
energy functional for the t'Hooft-Polyakov monopole in flat space-time. They
demonstrated the existence of a non-abelian solution, which asymptotically had
all the properties of the t'Hooft-Polyakov monopole. But the solutions they
found appear to have a naked singularity at the origin and the question
remained whether the singularity could be removed by an ansatz for the
background metric, that is more general than what was assumed or whether the
singularity was because of the Higgs field being frozen at a critical value
throughout space. In the present work, we address some of these issues.

In the next section, we set up our general formalism. In Section III, after a
brief review of the energy functional in the case of monopole in flat
space-time, we consider the same problem in curved space-time. Using a
Bogomol'nyi-type analysis, we find a positive-definite energy functional with a
lower bound. We discuss how the total energy and mass of the monopole are
related to the asymptotic behaviour of the metric functions and the Higgs
fields on the one hand and the topological winding number on the other hand.
Two extreme cases for a gravitating monopole arise. In one case, the Higgs
field is frozen to its vacuum expectation value (VEV) throughout space. The
mass of the monopole is then determined by its magnetic charge and the
magnitude of
the VEV. In the other case, in the comoving frame, the topological charge is
related to the asymptotic behaviour of the Higgs field. In Section IV, starting
from the most general static, spherically symmetric form for the metric we
calculate the curvature coefficients, components of energy-momentum tensor and
write the coupled equations for
Einstein, Yang-Mills and Higgs fields. In Section V, we specialize to the
isotropic and standard coordinate systems and arrive at the non-linear
differential equations whose solutions determine the solutions to the entire
system of equations. We discuss the numerical solutions, their main features
and their physical implications. The final section is devoted to summary.
\section{General framework}
We begin by considering a modified EYMH system, defined by the action $S$,
where
\begin{equation}
S=\int d^4x \sqrt{-g}\Big[ -{1 \over 16\pi Gv^2} R \Phi ^2 -{1\over 4}
(F_{\mu \nu}^a)^2 + {1\over 2}({\cal D }_\mu \Phi^a)^2 -{\lambda \over
4}(\Phi^2-v^2)^2\Big]
\end{equation}
In our conventions, $g_{\mu \nu}$ has the signature $(+ - - - )$. Indices $\mu,
\nu,$ ... run from 0 to 3 and $i,j,...$ from 1 to 3. Indices $a,b,...$ run over
the adjoint representation of a given gauge group. $R$ is the Ricci scalar.
$F_{\mu\nu}^a$ is the field strength associated with the gauge field $A_\mu^a$
\begin{equation}
F_{\mu \nu}^a= \partial_\mu A_\nu^a - \partial_\nu^a + g f^{abc} A_\mu^b
A_\nu^c  ,
\end{equation}
where g is the gauge coupling constant (not to be confused with the determinant
of the metric) and $f^{abc}$'s are the structure constants of the group.
The gauge-covariant derivative of the Higgs fiel $\Phi^a$ is given as follows
\begin{equation}
{\cal D}_\mu \Phi^a = \partial_\mu \Phi^a + g f^{abc} A_\mu^b\Phi^c
\end{equation}
In the broken phase of the gauge symmetry, when the Higgs field is frozen to be
$\Phi^2 = v^2$, the first term gives the usual Einstein Lagrangian.
The rest of the terms represent the standard Yang-Mills-Higgs system.
Here after, we shall work in units such that
\begin{equation}
4\pi Gv^2 = 1
\end{equation}
and then the action is
\begin{equation}
S = \int d^4x \sqrt{-g}\Big[ -{R\over 4}\Phi^2-{1\over
4}(F_{\mu\nu}^a)^2+{1\over 2}({\cal D}_\mu\Phi^a)^2 - {\lambda\over 4}
(\Phi^2-v^2)^2\Big]
\end{equation}
Varying the action (2.5) with respect to $ A^a_\mu, \Phi^a,$ and $ g_{\mu\nu}$
respectively we have the coupled equations for Yang-Mills and Higgs fields
\begin{equation}
{1\over\sqrt{-g}}{\cal D}_\mu\big(\sqrt{-g}F^{a\mu\nu}\big) =
- g f^{abc}\Phi^b{\cal D}^\nu\Phi^c  ,
\end{equation}

\begin{equation}
{1\over \sqrt{-g}}{\cal D}_\mu \big(\sqrt{-g}{\cal D}^\mu\Phi^a\big) =
- \Big({R\over 2}+\lambda(\Phi^2-v^2)\Big)\Phi^a  ,
\end{equation}
and the Einstein equation
\begin{equation}
G_{\mu\nu} = R_{\mu\nu} - {1\over 2} R = {2\over \Phi^2} T_{\mu\nu} ,
\end{equation}
where the energy-momentum tensor $T_{\mu\nu}$ is given by

\begin{eqnarray}
T_{\mu\nu}&  = &g_{\mu\nu}T - F^a_{\mu\rho}F^{a\rho}_\nu + {\cal
D}_\mu\Phi^a.{\cal
D}_\nu\Phi^a + \nabla_\mu \nabla_\nu\Phi^2\nonumber  \\
T & = & {1\over 4} F_{\mu\nu}^aF^{a\mu\nu} -{1\over 2}{\cal D}_\mu\Phi^a{\cal
D}^\mu\Phi^a + {\lambda \over 4}(\Phi^2 -v^2)^2 -{1\over 2}\Box\Phi^2
\end{eqnarray}

($\nabla_\mu$ is the space-time covariant derivative and the D'Alembertian
$\Box = {1\over\sqrt{-g}}\nabla_\mu(\sqrt{-g}\nabla^\mu)$.
The last terms in $T_{\mu\nu}$ and $T$ arise because of the coupling $R\Phi^2$
in the action [10].
\section{Static equations and Bogomol'nyi conditions}
We are interested in the static solutions of the system described in the
previous section.
Let us recall first the case of flat space-time and the well-known technique
due to Bogomol'nyi to find the static solutions of the system by solving
a set of first order differential equations called Bogomol'nyi equations [9].
Specializing to the t'Hooft-Polyakov monopole [11] with the gauge group
$SU(2)$, we know that the finite-energy configuration that describes it has a
finite extension outside of which we have the Higgs vacuum and hence the
condition
\begin{equation}
{\cal D}\Phi = 0 ,\hskip 12pt \Phi^2 = v^2.
\end{equation}
The gauge group is broken to the abelian $U(1)$. The gauge field that satisfied
(3.1) is given by [12]
\begin{equation}
A^a_i = {1\over gv^2}\epsilon^{abc}\Phi^b\partial_i\Phi^c + {1\over
v}\Phi^a A_i ,
\end{equation}
where $A_i$ is arbitrary and the corresponding $F_{ij}^a$ satisfies
\begin{equation}
\Phi^aF_{ij}^a = {\Phi^2\over v}{\cal F}_{ij},
\end{equation}
where
\begin{equation}
{\cal F}_{ij}= {1\over gv\Phi^2}
\epsilon^{abc}\Phi^a\partial_i\Phi^b\partial_j\Phi^c +\partial_i A_j
-\partial_j A_i .
\end{equation}
Only the above abelian static gauge field that clearly satisfies Maxwell's
equations will survive at large distances. Therefore,
we can define the "magnetic field" ${\cal B}_i$ due to the monopole at large
distances as
\begin{equation}
{\cal B}^i = {1\over 2} \epsilon^{ijk}{\cal F}_{jk} ={1\over
2gv^3}\epsilon^{ijk}\epsilon^{abc}\Phi^a \partial_j\Phi^b\partial_k\Phi^c +
\epsilon^{ijk}\partial_j A_k .
\end{equation}
Then the magnetic charge of the configuration is given by
\begin{equation}
{\cal G} = {1\over 4\pi}\int d^3x\partial_i {\cal B}^i =
{1\over 8\pi v^3g}
\int_{S^2_\infty} d\sigma_i\epsilon^{ijk}\epsilon^{abc}\Phi^a\partial_j\Phi^b
\partial_k\Phi^c = {n\over g} ,
\end{equation}
where n is the winding number of any maps $\Phi : S^2 \rightarrow S^2 $
defined as the surface integral [12]
\begin{equation}
n= {1\over 8\pi
v^3}\int_{S^2_\infty}d\sigma_i\epsilon^{ijk} \epsilon^{abc}\Phi^a\partial_j
\Phi^b\partial_k\Phi^c .
\end{equation}

Now, in
the limit ${\lambda \over g}\longrightarrow 0$, the energy functional for the
flat space-time is
\begin{equation}
{\cal E}=\int d^3x\Big[ {1\over 4}F_{ij}^aF^{aij}-{1\over 2}{\cal D}_i
\Phi^a {\cal D}^i\Phi^a \Big],
\end{equation}
and in terms of the "electric" and the "magnetic" fields,

\begin{eqnarray}
 E^a_i & = & {\cal D}_i \Phi^a ,\nonumber \\
 B_i^a & = & {1 \over 2} \epsilon _{ijk} F^{aij}  ,
\end{eqnarray}
the following inequality follows:

\begin{eqnarray}
{\cal E}& = &{1\over 2}\int d^3x \Big[ (E^a_i)^2 + (B^a_i)^2 \Big] =
\nonumber\\
& = & {1\over 2}\int d^3x(E^a_i \mp B^a_i)(E^a_i \mp B^a_i) \pm \int d^3x
E^a_i B^a_i \geq \pm \int d^3x E^a_i B^a_i .
\end{eqnarray}

Further, using the Bianchi identity for $F^a_{ij}$, we can express
\begin{equation}
\pm \int d^3x E^a_i B^a_i = \pm \int d^3x \partial_i\big({1\over 2}
\epsilon^{ijk} F_{jk}^a \Phi^a\big) = \pm \int_{S^2_R} d\sigma_i\big({1\over 2}
\epsilon^{ijk} F_{jk}^a\Phi^a\big).
\end{equation}
Comparing (3.11) and (3.3), we note that the surface integral in (3.11) becomes
asymptotically
\begin{equation}
\pm v\int d^3x \partial_i{\cal B}^i = {4\pi n v\over g}.
\end{equation}
On the other hand, if the Bogomol'nyi equations
\begin{equation}
E^a_i = \pm B^a_i
\end{equation}
are satisfied, the bound is saturated and
\begin{equation}
\pm \int_{S^2_R} d\sigma_i\big({1\over 2}\epsilon^{ijk}F_{jk}^a \Phi^a\big) =
\pm \int_{S^2_R} d\sigma_i \big({\cal D}^i\Phi^a\big)\Phi^a = \pm {1\over
2}\int_{S^2_R}d\sigma_i\partial^i(\Phi^2) .
\end{equation}
Therefore, if $\Phi^a \sim \big[ v -{\beta\over r} + {\cal O}({1\over r^2})
\big] $ as $ r\longrightarrow \infty$,
\begin{equation}
\pm {1\over 2}\int_{S^2}d\sigma_i\partial^i(\Phi^2) = \pm 4\pi v \beta +
{\cal O}\big({1\over r}\big),
\end{equation}
and from (3.12), (3.14), (3.15), we find
\begin{equation}
\beta = n .
\end{equation}
Thus, when Bogomol'nyi conditions are satisfied, we find a connection between
the asymptotic behavior of $\Phi$ and the topologically conserved number $n$.
Returning to curved space-time, we look for static solutions of equations
(2.6)-(2-8), in the
$A_0^a = 0 $ gauge. Since
\begin{equation}
F^a_{0i} = 0  ,\hskip 12pt {\cal  D}_0\Phi^a = 0 ,
\end{equation}
we find that the Eqs.(2.6) and (2.7) reduce to

\begin{eqnarray}
{1 \over \sqrt{-g}}{\cal D}_i \Big(\sqrt{-g} F^{a ij}\Big)& = &- g
f^{abc}\Phi^b
{\cal D}^j\Phi^c ,\\
{1\over\sqrt{-g}}{\cal D}_i\Big(\sqrt{-g}{\cal D}^i\Phi^a\Big)& = &
-\Big({R\over 2} + \lambda \big(\Phi^2 - v^2\big)\Big)\Phi^a .
\end{eqnarray}
The static energy functional that follows from (2.5) in curved space-time is
given by [8]
\begin{equation}
{\cal E} = \int d^3x \sqrt{-g}\Big[ {R\Phi^2 \over 4} +{1\over 4} F_{ij}^a
F^{aij} -{1\over 2}{\cal D}_i\Phi^a {\cal D}^i\Phi^a
+{\lambda \over 4}(\Phi^2-v^2)^2\Big]
\end{equation}
Clearly we need to modify the definitions (3.9) in flat space-time to
reduce (3.20) to a desired form analogous to (3.10). For this purpose following
Comtet, Forg\'acs and Horv\'athy [13], let us assume
\begin{equation}
{\cal D}_i\Phi^a + \partial_i u \Phi^a = {1\over 2} \sqrt{-\tilde
g}\epsilon_{ijk} F^{ajk},
\end{equation}
where u is an arbitrary time-independent function and $\tilde g = det|g_{ij}|$.
Equivalently,
\begin{equation}
{1\over 2}
\epsilon^{ijk} F_{jk}^a = \sqrt{-\tilde g}\big({\cal D}^i\Phi^a + \partial^iu
\Phi^a\big).
\end{equation}
Substituting (3.22) in (3.18) and using the Bianchi identity for $F^{a ij}$, we
find that (3.18) is satisfied provided
\begin{equation}
\partial_i u = {\partial_i\sqrt{g_{00}}\over \sqrt{g_{00}}}.
\end{equation}
With (3.23), after some algebra, the Higgs field Eq.(3.19) yields
\begin{equation}
{1\over\sqrt{-\tilde g}}\partial_i\Big[\sqrt{-\tilde
g}{\partial^i\sqrt{g_{00}}\over\sqrt{g_{00}}}\Big] = {R\over 2} + \lambda
(\Phi^2 - v^2).
\end{equation}
Comtet et al [13] have suggested that the Bogomol'nyi Eq.(3.21)
is the duality condition of an Euclidean Yang-Mills theory as it is in the flat
space time. Additionally, we shall show that this condition also saturates the
energy-functional.
Let us define the "electric" and "magnetic" fields as follows,

\begin{eqnarray}
E^a_i & = & ^4\sqrt{-g}\sqrt{-g^{ii}}\big({\cal D}_i\Phi^a +
{\partial_i\sqrt{g_{00}}\over \sqrt{g_{00}}}\Phi^a\big)=
^4\sqrt{-g}{\sqrt{-g^{ii}}\over g_{00}}{\cal D}_i\chi^a,\nonumber\\
B^a_i & = &^4\sqrt{-g}\sqrt{g^{jj}g^{kk}}{1\over 2}\epsilon_{ijk}F_{jk}^a,
\end{eqnarray}
where $\chi^a = \sqrt{g_{00}}\Phi^a$. The Bogomon'nyi conditions becomes
\begin{equation}
{\cal D}_i\chi^a = \sqrt{g_{00}}{\epsilon^{ijk}\over \sqrt{-\tilde g}}F_{ij}^a
{}.
\end{equation}
Using the Eqs.(3.24) and (3.25), we can express the energy-functional
${\cal E}$ as
\begin{eqnarray}
{\cal E} & = &{1\over 2}\int d^3x\Big[ (B^a_i)^2 + (E^a_i)^2 + {\sqrt{-g}\over
\sqrt{-\tilde g}}\partial_i\big(\sqrt{-\tilde g}{\partial^i\sqrt{g_{00}}\over
\sqrt{g_{00}}}\big)\Phi^2 - \lambda \sqrt{-g}(\Phi^2 - v^2)\Phi^2
\nonumber \\
& + &  {\lambda\over 2}
\sqrt{-g}(\Phi^2-v^2)^2 + \sqrt{-g}{\partial_i\sqrt{g_{00}}\over
\sqrt{g_{00}}}{\partial^i \sqrt{g_{00}}\over \sqrt{g_{00}}}\Phi^2
 + \sqrt{-g}{\partial^i\sqrt{g_{00}}\over\sqrt{g_{00}}}\partial_i
\Phi^2 ]   \nonumber\\
& = & {1\over 2}\int d^3x\Big[(B^a_i)^2 + (E^a_i)^2 +\partial_i \big(
\sqrt{-\tilde g}\partial^i\sqrt{g_{00}}\Phi^2\big) -
{\lambda \over 2} \sqrt{-g}(\Phi^4 -v^4)\Big] .
\end{eqnarray}
Ignoring the $\lambda $-term we have then
\begin{eqnarray}
{\cal E} & = &{1\over 2}\int d^3x \Big[ (B^a_i)^2 + (E_i^a)^2 \big]+\partial_i
\Big(\sqrt{-\tilde g}\partial^i\sqrt{g_{00}}\Phi^2\Big)]  \nonumber \\
& = & {1\over 2}\int d^3x \Big[\big( E^a_i\mp B^a_i\big)\big(E^a_i\mp
B^a_i\big)\Big] + \int d^3x \Big[\pm E^a_i B^a_i+ {1\over
2}\partial_i\big(\sqrt{-\tilde g}\partial^i\sqrt{g_{00}}\Phi^2
\big)] ,
\end{eqnarray}
and hence
\begin{equation}
{\cal E} \geq  \int d^3x \Big[\pm E^a_i B^a_i + {1\over
2}\partial_i\Big(\sqrt{-\tilde g}\partial^i\sqrt{g_{00}}\Phi^2\Big)].
\end{equation}
When the Bogomol'nyi conditions $ E^a_i = \pm B_i^a $ are satisfied, the
energy- functional reaches the lower bound and we have
\begin{equation}
{\cal E} = \int d^3x \Big[{-\sqrt{-\tilde g}\over \sqrt{g_{00}}}{\cal
D}_i\chi^a
{\cal D}^i\chi^a + {1\over 2}\partial_i\big(\sqrt{-\tilde
g}\partial^i\sqrt{g_{00}}.\Phi^2\big)] .
\end{equation}
Using (3.22), (3.23) and the Bianchi identity for $ F^{aij}$, we can derive the
 relation
\begin{equation}
{\cal D}_i\chi^a.{\cal D}^i\chi^a = {1\over 2}{\sqrt{g_{00}}\over \sqrt{-\tilde
g}}.\partial_i\Big({\sqrt{-\tilde g}\over \sqrt{g_{00}}}\partial^i\chi^2\Big),
\end{equation}
and hence
\begin{equation}
{\cal E} =  {1\over 2}\int d^3x \partial_i\Big[ \sqrt{- g}\big(-{1\over
2}{\partial^i g_{00}\over g_{00}}\Phi^2 - \partial^i\Phi^2\big)\Big]
 = {1\over 2}\int_{S^2_R}d\sigma_i\Big[-{1\over 2}{\partial^ig_{00}\over
g_{00}}\Phi^2-\partial^i\Phi^2\Big]\sqrt{-g} .
\end{equation}
Now, if we have a finite-energy configuration with finite extension and the
space-time is asymptotically flat, we must have
\begin{equation}
{\cal D}\chi = 0,\hskip 12pt \chi^2 = v^2,
\end{equation}
and an exact analogy prevails with the flat space-time t'Hooft-Polyakov
monopole
considered before. We have asymptotically an abelian magnetic field given by
\begin{equation}
{\cal B}^i =  {1\over 2}{\epsilon^{ijk}\over \sqrt{-\tilde g}}{\cal F}_{jk},
\end{equation}
where
\begin{equation}
  {\cal F}_{jk} ={1\over 2v^3g} \epsilon^{abc}\chi^a \partial_j \chi^b
\partial_k\chi^c +\partial_j A_k-\partial_k A_j ,
\end{equation}
with an arbitrary $A_j$ and as before. The magnetic charge ${\cal G}$ is given
by
\begin{equation}
4\pi{\cal G} = \int d^3x\partial_i {\cal B}^i = {2\pi\over
v^3g}\int_{S^2_\infty}\epsilon^{abc}\epsilon^{ijk}\chi^a\partial_j\chi^b
\partial_k\chi^c d\sigma_i =  4\pi{n\over g},
\end{equation}
where n is the winding number of the mapping $\chi : S^2_\infty \longrightarrow
S^2 $, and
\begin{equation}
n = {1\over 8\pi v^3}\int_{S^2_\infty}\epsilon^{abc}\epsilon^{ijk}\chi^a
\partial_j\chi^b\partial_k\chi^c d\sigma_i .
\end{equation}
As before, we can show that, asymptotically,
\begin{equation}
{\cal E} = \int d^3x \partial_i {\cal B}^i = {4\pi n \over g}
\end{equation}
However, if as  $r \longrightarrow \infty $,
\begin{eqnarray}
 g_{ij}& \longrightarrow & 1 + {\cal O}\big({1\over r}\big),\nonumber\\
 g_{00}& \longrightarrow & 1 -{M\over 2\pi v^2 r}
+{\cal O}\big({1\over r^2}\big),\nonumber\\
\Phi^2 & \longrightarrow & v^2 -{2v\beta \over r} +{\cal O}\big({1\over
r^2}\big),
\end{eqnarray}
where $M$ is defined as the monopole's mass, then from (3.32), we have
\begin{equation}
{\cal E} = {1\over v}M + {4\pi\over g}\beta .
\end{equation}
{}From (3.38) and (3.40), we find that
\begin{equation}
n = {1\over 4\pi v}Mg + \beta .
\end{equation}
We have two extreme cases for a gravitating monopole. In the first case, the
Higgs field is frozen and the mass of the monopole is determined by the
magnetic charge and the value of $v$
\begin{equation}
M= n. {4\pi\over g} v = 4\pi v {\cal G}.
\end{equation}
 As the topological charge is invariant, the mass of our monopole is also
conserved. It indicates that in this case our monopole is not a dissipative
one. In the second case, we are in the comoving frame of reference, in which
case
\begin{equation}
 g_{00} = 1  \hskip 12pt \text{and hence} \hskip 12pt n = \beta,
\end{equation}
and thus the topological charge is determined by the asymptotic behavior of the
 Higgs field only.
\section{Spherically symmetric ansatz and basic equations}
We define a spherically symmetric static metric by
\begin{equation}
ds^2 = A^2(r) dt^2 - B^2(r) dr^2 - C^2(r) r^2 d\Omega .
\end{equation}
In fact we could always redefine the "radius" $r$ so that the line-element
could be represented by two metric functions only. But we prefer the above more
general form since we intend to consider both the standard and the isotropic
coordinate systems. It is more convenient for our purposes to have the more
general metric form (4.1) and specialize to the chosen coordinate system by
suitably identifying the functions.

The Einstein tensor and the Ricci scalar curvature are then easily computed.
\begin{eqnarray}
G_{00}& = & {A^2\over B^2}\Big[{1\over r^2}\big({B^2\over C^2}-1\big) +
{2\over r}{B'\over B} + 2 {B'\over B}{C'\over C}
 -  2{C''\over C} - {6\over r}{C'\over C} - \Big({C'\over C}\Big)^2\Big],
\nonumber \\
G_{rr}& = & \Big({C'\over C}\Big)^2 + {2\over r}{C'\over C} + {1\over r^2}
\Big(1 - {B^2\over C^2}\Big) + 2{C'\over C}{A'\over A} +
{2\over r}{A'\over A}, \nonumber\\
G_{\theta \theta } & = & {C^2 r^2\over B^2}\Big({A''\over A} + {1\over r}{ A'
\over A} - {A'\over A}{B'\over B} + {C'\over C}{A'\over A} - {1\over r}{B'\over
B} - {B'\over B}{C'\over C} + {C''\over C} + {2\over r}{C'\over C}\Big),
\nonumber \\
G_{\phi \phi } & = & \sin^2\theta \,G_{\theta \theta },
\end{eqnarray}
and
\begin{eqnarray}
R & = & {2\over r^2}\Big({1\over B^2} - {1\over C^2}\Big) + {2\over B^2}\Big[
{A''\over A} + {2\over r}{A'\over A} -{B'\over B}{A'\over A} -
{2\over r}{B'\over B} - 2 {B'\over B}{C'\over C} + 2 {C'\over C}{A'\over A}
\nonumber \\
  & & + 2{C''\over C} + {6\over r}{C'\over C} + \Big({C'\over C}\Big)\Big] .
\end{eqnarray}
The Bogomol'nyi equations (3.21) along with (3.23) expressed in spherical
polar coordinates take the form
\begin{eqnarray}
{\cal D}_r\Phi^a + {A'\over A}\Phi^a & = & - {B\over C^2 r^2 \sin \theta }
F^a_{\theta \phi } ,\\
{\cal D}_\theta \Phi^a & = & - {1\over B \sin \theta } F^a_{\theta r}, \\
{\cal D}_\phi\Phi^a  & = & - {\sin \theta \over B} F^a_{r \theta}.
\end{eqnarray}
The 'Hooft-Polyakov spherically symmetric ansatz for $\Phi $ and $A_\mu^a $,
\begin{eqnarray}
\Phi^a & = & v h(r) r^a, \\
A_i^a(r)& = & \epsilon _{iab} {\hat r^b \over gr}\big( 1 - W(r) \big) ,
\end{eqnarray}
transformed into spherical polar coordinates lead to
\begin{eqnarray}
\Phi^a & = & v h(r) (\sin \theta \cos \phi, \sin \theta \sin \phi, \cos \theta
) ,\\
F^a_{r \theta } & = & - {W'\over g}\big( \sin\phi, - \cos\phi , 0 \big)
, \\
F^a_{r \phi } & = & - {W'\over g}\big( \cos \phi \cos \theta, \sin\phi
\cos\theta , -\sin \theta \big)\sin\theta ,\\
F^a_{\theta \phi } & = & {W^2 - 1 \over g}\big(\cos\phi \sin\theta, \sin\phi
\sin\theta, \cos\theta \big)\sin\theta ,
\end{eqnarray}
and hence
\begin{eqnarray}
{\cal D}_r\Phi^a & = & v h'(r)\hat r^a ,\nonumber \\
{\cal D}_\theta \Phi^a & = & v h W \big(\cos\theta, \cos\theta \sin\phi, -\sin
\theta \big),\nonumber \\
{\cal D}_\phi \Phi^a & = & - W h v \big(\sin\theta \sin\phi, -\sin\phi
\cos\phi , 0\big) .
\end{eqnarray}
Substituting (4.13) into (4.4)-(4.6) we obtain two independent equations,
\begin{eqnarray}
h' + {A'\over A} h & = & {B\over C^2 r^2}{1- W^2\over v g},\nonumber \\
W'& = & - v g B h W .
\end{eqnarray}
Likewise, the components of the energy-momentum tensor are calculated to be
\begin{eqnarray}
T_{00} & = & {\Phi^2 A^2\over B^2}\Big[2{C'\over C}{A'\over A} + {2\over
r} {A'\over A} - {A'\over A}{B'\over B} + {A''\over A} - {1\over
2}\big({A'\over A}\big)^2 + {\lambda v^2\over 4}{(h^2 - 1)^2\over h^2} +
\nonumber\\
& &+ 2{h''\over h} + 2\big({h'\over h}\big)^2 + 2 {h'\over h}\big(2{C'\over C}
+ {2\over r} - {B'\over B}\big) + 3 {A'\over A}{h'\over h}\Big]\nonumber\\
T_{rr} & = & \Phi^2\Big[ - {1\over 2}\big({A'\over A}\big)^2 -{A'\over A}
{h'\over h} + 2{B'\over B}{h'\over h} -{\lambda v^2\over 4} B^2 {(h^2-1)^2
\over h^2}\Big]\nonumber\\
T_{\theta\theta}&= &{C^2 r^2\over B^2}\Phi^2\Big[{1\over 2}\big({A'\over A}
\big)^2
-{h''\over h} -\Big({h'\over h}\Big)^2-{h'\over h}\Big(2{C'\over C}+{2\over r}
-{B'\over B}\Big) - {\lambda v^2\over 4} B^2{(h^2 - 1 )^2\over h^2}
\Big]\nonumber\\
T_{\phi\phi} & = & \sin^2\theta T_{\theta\theta}
\end{eqnarray}
Then (4.2), (4.3) and (4.15) yield Einstein's equations in the static case with
the ansatz (4.1). Rather than writing these equations explicitly, we give below
the equivalent set of four independent equations that follow from Einstein's
equations and the Eq.(3.24)
\begin{eqnarray}
4{h''\over h} + 4 \Big({h'\over h}\Big)^2 + 4{h'\over h}\Big(2{C'\over C} +
{2\over r}-{B'\over B}\Big) + 6{h'\over h}{A'\over A} & = & {\lambda v^2\over
2}B^2 {(h^4 - 1)\over h^2} ,\\
4{h'\over h}\Big({B'\over B}+{A'\over A}\Big)- \lambda v^2 B^2 {(h^2 - 1)^2
\over h^2} & = & 0,\\
\Big({C'\over C} + {1\over r} + {A'\over A}\Big)^2 + 6 {h'\over h}{A'\over A}
 - {B^2\over C^2 r^2} & = & {\lambda v^2\over 2}B^2 {(h^2-1)^2\over h^2} ,\\
\Big({C'\over C} + {1\over r} + {A'\over A}\Big)'+\Big({C'\over C}+{1\over r}
- {B'\over B}\Big)\Big({C'\over C}+{1\over r}+{A'\over A}\Big)& = &
3 {h'\over h}{A'\over A}-{\lambda v^2\over 4} B^2{(h^2-1)(3h^2-1)\over h^2} ,
\end{eqnarray}

This set of equations together with the Bogomol'nyi equations (4.14) are the
basic equations of our system.

{ \it a) Higgs vacuum:} \\
The above set of equation simplify greatly if Higgs field is frozen in
the broken phase $h^2=1$.
The Eqs.(4.16) and (4.17) are satisfied automatically. The equation
(4.19) can be firstly integrated. It coincides exactly with the equation
(4.18), if the integration constant is chosen as $\pm 1$. So the set of
equations to be solved reduces to
\begin{eqnarray}
{C'\over C} + {1\over r} + {A'\over A} & = & \pm {B\over Cr }\\
{A'\over A} & = & {B\over C^2 r^2}{1 - W^2\over v g}\\
W' & = & - v g B W
\end{eqnarray}
The number of unknown functions can be reduced to three as we still have the
freedom to fix the scale of radius using diffrent cordinate systems.
 We note that, the mass of the monopole is determined
by the formula (3.42) with a given topological charge and the vacuum
expectation value $v$ of the Higgs field. The surviving abelian magnetic
field at large distances is determined completely in terms of the metric
functions according to the Eq.(3.34)
\begin{equation}
B^r={1\over 2 g} {A^3\over BC^2r^2}
\end{equation}
{\it b) Isotropic coordinate system}\\
In the isotropic coordinate system we choose the coordinate fixing condition
\begin{equation}
B = C ={\tilde B\over A}.
\end{equation}
The line-element in this coordinate system takes the form
\begin{equation}
ds^2= A^2(r) - {\tilde B^2(r)\over A^2(r)}\Big( dr^2 + r^2 d\Omega\Big).
\end{equation}
Instead of the variable $r$, let us introduce the new variable $x=v g r$.
The system of equation (4.20)-(4.22) then reduces to
\begin{eqnarray}
{\tilde B'(x)\over \tilde B(x)} + {1\over x} & = & \pm {1\over x},\\
{A'(x)\over A^2(x)} & = & {1-W^2(x)\over \tilde B(x)x^2},\\
W'(x) & = & - {\tilde B(x)\over A(x)} W(x),
\end{eqnarray}
where prime now denotes differentiation with respect to $x$.

The Eq.(4.26) gives two possibilities i) With positive
sign on the right-hand side $\tilde B = constant$, which can be chosen to be
unity without any loss of generality and ii)$\tilde B = {const \over x^2}$ with
negative sign on the right-hand side.

The first possibility has been studied by Balakrishna and Wali [8]. Their
abelian solution with $ W = 0 $ gives
\begin{equation}
A =  1 - {1\over x+1}
\end{equation}
with the asymptotic condition $ A\longrightarrow 1$ as $ x\longrightarrow
\infty $. It is in reality Majumdar-Papapetrou [3] solution representing a
magnetic monopole with its mass equal to its magnetic charge or equivalently an
extreme Reissner-Nordstr\"om blackhole.

The non-abelian solution corresponding to this case involves the system of
non-linear differential equations
\begin{equation}
r^4F'' =e^{-2F} , \hskip 12pt {g\over V} ={1\over r} + F',
\end{equation}
where $ W = e^{-F}/ r $. For the discussion of the solutions to (4.30) and
their implications, the reader is referred to ref.[8].

The second possibility leads to the following equations
\begin{eqnarray}
{A'\over A^2} & = & {1-W^2\over \alpha}\nonumber\\
W' & = & - \alpha {W\over x^2 A},
\end{eqnarray}

Let us define $W=x e^{-F} $. Then, we have the non-linear system
\begin{eqnarray}
F'' + {2\over x} F' & = & e^{-2F}\\
A & = & {\alpha \over x(x F'-1)}.
\end{eqnarray}
The Eq.(4.32) is well-known in mathematical physics as the classical
Poisson-Boltzmann differential equation that plays a central role in the
description of the thermal exploding matter [14].
The space-time is not asymptotically flat in the ordinary manner, which can be
seen in the following way.

If we make the transformation $ z={1\over x}$, the Eq.(4.32) takes the
form,
\begin{equation}
z^4{d^2F\over dz^2} = e^{-2F},\hskip 12pt A=-{\alpha z \over (z{dF\over dz}+1)}
\end{equation}
which is again the Eq.(4.30), which belongs to the class of Lane-Emden
equations. As $ z \longrightarrow 0 $ (or $ x\longrightarrow \infty $), we know
[8]
\begin{equation}
F(z)\longrightarrow -lnz + \gamma \sqrt{z} \sin\Big[{\sqrt{7}\over 2}ln z
+\delta\Big],
\end{equation}
and hence,
\begin{equation}
A\longrightarrow {- 2\alpha \over \gamma}{\sqrt{z}\over \sin\big[{\sqrt{7}\over
2} lnz + \delta\big] + \cos \big[{\sqrt{7}\over 2}lnz +\delta \big]}
\longrightarrow 0
\end{equation}
{\it c) Standard coordinate system}\\
In the standard coordinate system, the coordinate fixing condition is
\begin{equation}
C = 1
\end{equation}
Again with the variable x, the equations to be solved are
\begin{eqnarray}
{A'\over A} + {1\over x} & = & \pm {B\over x}\\
{A'\over A} & = & {B\over x^2}\big(1-W^2\big)\\
W' & = & - B W
\end{eqnarray}
We can prove that the case with minus sign in (4.38) contradicts the condition
on asymptotic behavior. Introducing the new variable $ y= W^2 $, after a
little algebra,
the case with positive sign leads us to the system of equations
\begin{eqnarray}
y' & = & - {2x y\over y-1 + x},  \\
B  & = & {x \over y-1 + x },     \\
{A'\over A} & = & {1-y\over x(y+x-1)} .
\end{eqnarray}
The Eq.(4.41) is well-known in mathematical physics as Abel's
differential equation of the second type. The Abel equation has clearly a
solution $ y = 0$. In this case we have an Abelian magnetic monopole which at
the same time is an extreme Reissner-Nordstr\"om blackhole
\begin{eqnarray}
B  & = & (1-{1\over x})^{-1} \\
A  & = & (1-{1\over x})
\end{eqnarray}
This monopole has mass $ M=4\pi v/g $.

The Abel equation (4.41) has no known analytic solution other than $y=0$.
Before proceeding
to solve it numerically, let us attempt to understand the nature of the
solutions by a qualitative analysis of the equation [15]. Considering $x(t)$
and $y(t)$ as functions of a parameter $t$ we can write the Abel equation as a
pair of equations for $x(t)$ and $y(t)$ as
\begin{eqnarray}
{dy(t)\over dt}  & = & -2x(t)y(t)\\
{dx(t)\over dt}  & = & y(t) + x(t) -1
\end{eqnarray}

This pair has two finite singular points in the $(x,y)$ plane at $(0,1)$
and $(1,0)$. The characteristic matrix at the point $(0,1)$ has complex
eigenvalues. It is described as the focus and it can be proved that the
solution curves spiral around and converge to this point.
At the other point, the characteristic matrix has real eigenvalues. Such a
point is described as the saddle point. The solution curves (except the curves
known as separatrices) approach the saddle point and then go away from it.
The separatrices in our case is the curve passing through $(0,y_{0c})$ and
$(1,0)$ and the curve $y=0$. These features are illustrated in Fig.1.

In Fig.2, we plot the family of solutions of (4.41). Consistent with the above
qualitative analysis, we find, with the initial value $y_0$ greater than a
critical value $y_{0c} =1.661$, the curves $ y(x) = W^2(r)$ fall of
exponentially for large $r$. With the initial value smaller
than the critical value, the curves spread only to a finite value of $r$ and
return to the y-axis as parts of spiral curves in the positive quadrant of the
$(x,y)$ plane. With $y_0=y_{0c}$, the curve passes the x-axis at $ x=1$
and then continues on to negative values. Since $ y =W^2 $, we are not
interested in negative values of $y$.

Since we are interested in monopole solutions, let us first concentrate our
attention on the family of
solutions for $W^2$ that fall off exponentially.  In Fig.3 we plot
such a solution along with the metric functions obtained by solving (4.42) and
(4.43). With the aid of (4.41) we can prove that $B$ is regular and $A$ has no
zeroes; it has a
singularity at the origin. We recall from (4.10)-(4.12) that the monopole has
a nontrivial, non-abelian magnetic core when $W$ does not vanish.
Thus we have a monopole with a non-abelian core. The singularity at the origin
is not hidden by a blackhole horizon as in the case of
Majumdar-Papapetrou monopole. However, it is worth noting that the metric
function $A$ has a minimum inside the core of the monopole. By changing the
value of the integration constant of the Eq.(4.43),
this minimum can be made arbitrarily close to zero with a corresponding
increase in the height of $B$ (Fig.4). But by analytic arguments we can show
that for the exponentially falling solutions of $y$, the metric function $A$
which satisfies the asymptotic condition $A\longrightarrow 1$ as $
x\longrightarrow \infty$ cannot strictly vanish
at any finite $x$. This raises the following important question: can
such a metric function which has a minimum arbitrarily close to zero be
interpreted as a blackhole horizon? In recent numerical studies, such
numerical minima are indeed interpreted as horizons.

Finally we come to the separatrice that passes throught $(0,y_{0c})$ and
$(1,0)$. Its continuation is $ y=0$ for $x\geq 1$. Taking this as a solution
for $y$, we see that (4.43) gives
\begin{equation}
A(x)= 1 - {1\over x} \hskip 12pt \text{for}\hskip 12pt x > 1,
\end{equation}
where we have chosen the constant of integration to be unity so that
$A(x)\longrightarrow 1$ as $x\longrightarrow \infty$ and $A(x)=0$ at $x=1$.
It is clear from (4.42) that $B(x)$ also satisfies the asymptotic condition,
It is infinite at $x=1$.(See Fig.5) and is regular for $x < 1$. Now for $x\leq
1 , A = 0$ is a solution. Further, we can verify that the Ricci scalar $ R$
curvature is finite at $x \leq 1 $ and diverges only at $x=0$. We have,
therefore, a solution with a blackhole horizon at $x=1$, surrounding the
singularity at the origin. Inside the horizon, we have nonabelian field
components since $W^2\not=0$. Thus, the solution represents the exact analog
and genaralization of the Majumdar-Papapetrou solution in the case of
Einstein-Maxwell system.
\section{Conclusion}
Extended objects like non-abelian monopoles provide a fertile ground for the
study of the interplay between gravitational and gauge interactions with a
spontaneously broken symmetry. Previous work by others and ours reported in
this paper show conclusively that the gravitational interactions, although
weak, can have significant effects on the properties of t'Hooft-Polyakov type
monopoles and therefore, cannot be ignored in quantum aspects of such extended
objects.

By considering an unconventional coupling of the Higgs field to gravity, we
have sought static sperically symmetric solutions to Einstein-Yang-Mills-Higgs
system. In contrast to the conventional studies of such a system with Einstein
action, our unconventional coupling is better suited for the use of the
Bogomol'nyi type analysis in reducing the solutions to the complex system of
coupled equations to the solution of a single non-linear equation whose
solution determines the solution of the entire system. We begin with the most
general spherically symmetric metric and specialize to isotropic and standard
coordinate systems. In the case of the former, we extend the previous study of
Balakrishna and Wali [8] to find a new type solution. The main results are in
the case of standard coordinate system. We find both abelian and non-abelian
type solutions when Higgs field is frozen to its vacuum expectation value.

The abelian solution presents a generalization of the Majumdar-Papapetrou
solution in the case of coupled Einstein-Maxwell system. The monopole in this
case is also an extreme Reissner-Nordstr\"om blackhole. In the non-abelian
case, the non-linear equation happens to be the well-known Abel equation of the
second type. We study the solutions of this equation numerically and find two
types of solutions. In one family of solutions, the non-abelian parts of the
gauge fields fall off exponentially as in the case of t'Hooft-Polyakov
monopoles in flat space-time. Asymptotically, we have the field of a monopole
with total charge equal to its mass. However, the solution has a singularity at
the origin which, strictly speaking, is not enclosed in a blackhole horizon.
The metric function $g_{00}$ has a minimum at a finite radius and can be made
arbitrarily small. The corresponding metric function $g_{rr}$ has a maximum.
These features suggest the presence of a horizon. However, from analytical
arguments, we can show that $g_{00}$ cannot vanish. This leaves the physical
interpretation of such a situation unclear and needs further study.

We do find a solution however, with a horizon at $vgr=x=1$. This solution has
only the Abelian component for $x\geq 1$ and behaves like a monopole for large
distances with a mass $ M=v4\pi /g$. It charge is ${\cal G}=1/g$. Remembering
that we are working in the units $4\pi Gv^2=1$, we can reexpress the mass as
$M=\sqrt{{4\pi \over G}}{1\over g}$. Similarly, the magnetic charge turns out
to
be $ \sqrt{4\pi }{\cal G} = {\cal Q} ={\sqrt{4\pi }\over g}$. Thus $GM^2={\cal
Q}^2$, and hence we have
an extreme Reissner-Nordstr\"om blackhole with magnetic charge equal to its
mass. It has the unusual feature such that the metric function $g_{00}$
vanishes
for $x\leq 1$.
 \section{Acknowledgment}
This work was supported in part by the U.S. Department of Energy under contract
number DE-FG02-85ER40231.

Nguyen Ai Viet is grateful to Professor A.Zichichi for a World Laboratory
scholarship. K.C.Wali would like to thank to Professor R.F.Sawyer for the
hospitality
at the Physics Department, University of California, Santa Barbara where this
work was partially done. He would also like to acknowledge Professor J.B.
Hartle for helpful discussions.

\newpage
\figure{ Fig.1  Singular points in the $(x,y)$ plane of the Abel equation's
solution curves. The point $(0,1)$ is a focus, while the one $(1,0)$ is a
saddle point with two bold dashed-dotted lines as its separatrices. Solutions
spiral around a focus; solutions approach, but do not pass the saddle points.}

\figure{ Fig.2  Solutions of the Abel equation $y' = -{2yx\over y-1+x}$.
Solutions with $y_0 > y_{0c}$, fall off exponentially. Solutions with $y_0 <
y_{0c}$ are shown with dotted lines. The solid line is the separatrice
solution}

\figure{ Fig.3  A monopole solution in curved space time. The solid and dotted
curves represent the metric functions $A$ and $B$ respectively. the
dashed-dotted
curve is $y=W^2$. The curves are plotted as function of $ln(1+x)$}.

\figure{ Fig.4  Monopole solutions with the minimum of the metric function $A$
varrying. The minimum can be made arbitrarily close to zero but never zero. We
can prove that if $A=0$ at some x it has to be zero everywhere. The solid,
dotted and dashed-dotted curves represent the metric functions $A$, $B$ and the
gauge field $y=W^2$ respectively. The curves are plotted as function of
$ln(1+x)$}

\figure{ Fig.5 A monopole-blackhole solution. The solid, dotted and
dashed-dotted curves represent the functions $A, B$ and $y=W^2$. The curves are
plotted as function of $ln(1+x)$. The non-abelian magnetic core
is inside the horizon at $x = 1 $. Outside of the core $W^2=0$ and we have an
abelian field. }
\end{document}